\documentclass[10pt,conference]{IEEEtran}
\usepackage{times}
\usepackage{helvet}
\usepackage{courier}
\usepackage{multirow}
\usepackage{bm}
\usepackage{amsmath}
\usepackage{amsfonts}
\usepackage{graphicx}
\usepackage{xcolor}
\usepackage{multirow}

\usepackage{url}

\newcommand{\nop}[1]{}

\newcommand{\Wendy}[1]{} 
\newcommand{\system}{DeepIDEA}

\title{Cyber Intrusion Detection by Using Deep Neural Networks with Attack-sharing Loss}
\author{\IEEEauthorblockN{Boxiang Dong\IEEEauthorrefmark{1},
Hui (Wendy) Wang\IEEEauthorrefmark{2}, Aparna S. Varde\IEEEauthorrefmark{1}, Dawei Li\IEEEauthorrefmark{1}, Bharath K. Samanthula\IEEEauthorrefmark{1}, \\Weifeng Sun\IEEEauthorrefmark{3}, Liang Zhao\IEEEauthorrefmark{3}}
\IEEEauthorblockA{\IEEEauthorrefmark{1}Montclair State University, Montclair, New Jersey 07043\\
Email: \{dongb, vardea, dawei.li, samanthulab\}@montclair.edu}
\IEEEauthorblockA{\IEEEauthorrefmark{2}Stevens Institute of Technology, Hoboken, New Jersey 07030\\
Email: \{Hui.Wang\}@stevens.edu}
\IEEEauthorblockA{\IEEEauthorrefmark{2}Dalian University of Technology, Dalian, China 116024\\
Email: \{wfsun, liangzhao\}@dlut.edu.cn}
}

\begin{document}
\maketitle

\begin{abstract}
Cyber attacks pose crucial threats to computer system security, and put digital treasuries at excessive risks. This leads to an urgent call for an effective intrusion detection system that can identify the intrusion attacks with high accuracy. It is challenging to classify the intrusion events due to the wide variety of attacks. Furthermore, in a normal network environment, a majority of the connections are initiated by benign behaviors. The class imbalance issue in intrusion detection forces the classifier to be biased toward the majority/benign class, thus leave many attack incidents undetected. 
Spurred by the success of deep neural networks in computer vision and natural language processing, in this paper, we design a new system named \system ~that takes full advantage of deep learning to enable intrusion detection and classification. 
To achieve high detection accuracy on imbalanced data, we design a novel attack-sharing loss function that can effectively move the decision boundary towards the attack classes and eliminates the bias towards the majority/benign class. 
By using this loss function, \system ~respects the fact that the intrusion mis-classification should receive higher penalty than the attack mis-classification.
Extensive experimental results on three benchmark datasets demonstrate the high detection accuracy of \system. In particular, compared with eight state-of-the-art approaches, \system ~always provides the best {\em class-balanced accuracy}.
\end{abstract}
\section{Introduction}
\label{sc:intro}
Recent years witness an expeditious outbreak of cyber attacks. Online Trust Alliance \cite{cyber} revealed that 2017 is ``the worst year ever'' in data breaches and cyber attacks around the world.
The amount of disclosed cyber incidents targeting businesses nearly doubled from 82,000 in 2016 to 159,700 in 2017. 
The penetration attack at Equifax leaked the financial credit report of 145 million consumers, which constitutes 45\% of the total population in the U.S. 
The WannaCry ransomware attack infected 300,000 computer systems within four days, and severely disrupted the medical appointments in the U.K..
These catastrophic attacks bring forth the most intensive aspirations for an effective intrusion detection system (IDS) that can identify the intrusion with high accuracy. 

Traditional signature-based IDS techniques heavily depend on the signature database constructed by security experts, and thus fail to detect novel attacks.
A wide variety of data mining and machine learning models, e.g., decision tree, support vector machine (SVM), and graph mining algorithms \cite{almseidin2017evaluation,biswas2018intrusion,dong2017efficient}, have been adapted to discover anomaly from the network monitoring data. However, they are not favorable at representing intrusion detection classification functions that have many complex variations \cite{bengio2009learning}. 

Recently, deep learning emerges as a favorable solution to dealing with complicated input-output mappings. Its application in computer vision and natural language processing leads to breakthroughs in these areas. 
Specifically, it builds a neural network by stacking a certain number of layers of neurons. With sufficiently large number of layers and units, a deep network can represent functions of high complexity. 
Compared with traditional machine learning models, it avoids the need for feature extraction. Most importantly, it produces the best-in-class accuracy by learning from a large amount of labeled data.

Quite a few latest research in intrusion detection resort to deep learning. Most of them \cite{javaid2016deep,chowdhury2017few} simply learn a new feature representation by using various deep neural networks (e.g., deep autoencoder and convolutional networks), and then rely on traditional classifiers such as SVM and k-nearest neighbor (KNN) to detect attacks. 
Kitsune \cite{mirsky2018kitsune} is the most recent work that detects network intrusion attacks with deep neural networks. It applies an ensemble of autoencoders to learn the identify function of the original data distribution. For any new instance, its anomaly score is calculated based on the distance between the autoencoders' output and its feature values. 
However, we argue that such a design only employs deep learning to discover inherent/generic features in network connections, but fails to take advantage of its capacity to learn complex classification functions. \Wendy{I don't fully understand this sentence. Do you mean the following: "However, Kistune only can discover inherent/generic features in network connections that can be used for intrusion detection. It does not support classification of intrusion attacks."}

\begin{figure}[!htbp]
    \centering
    \includegraphics[width=0.4\textwidth]{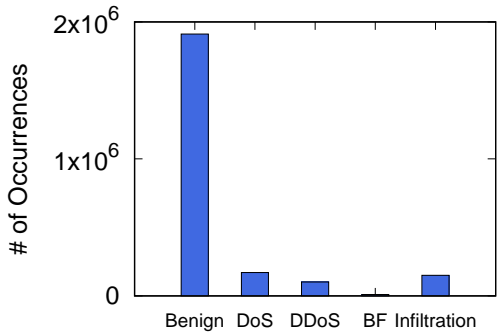}
    \vspace{-0.15in}
    \caption{Class distribution of the CICIDS17 dataset. There are five classes, including the benign event and four types of attacks, namely DoS, DDoS, brute-force (BF), and infiltration attacks. }
    \label{fig:dist}
    \vspace{-0.15in}
\end{figure}

There are two major challenges of designing a deep neural network based intrusion detection system. We discuss these two challenges below. 

\noindent{\bf Challenge 1: diversity of intrusion attacks.} There are many types of intrusion attacks, each exploiting a wide range of techniques to conduct the invasion. 
    Even the same type of attacks can exhibit different behavior patterns. 

\noindent{\bf Challenge 2: imbalanced class distribution.} In a healthy network environment, a majority of the connections are benign. This makes the network connection instances follow a long-tail class distribution. 
Moreover, different types of intrusion attacks are unevenly distributed in practice. As an example, consider a real-world network intrusion detection dataset named CICIDS17, which is collected and released by Canadian Institute for Cybersecurity. The data is labeled with 5 classes, including the benign class and four types of attacks, namely DoS, DDoS, brute-force (BF), and infiltration attacks. The distribution of the intrusion attacks is illustrated in Figure \ref{fig:dist}. Apparently, the classes are highly imbalanced.
The imbalanced class distribution forces the classifier models to be biased toward the majority class, and thus lead to poor accuracy on the minority classes (i.e., the intrusions). 

To address these two challenges, we build a new intrusion detection and classification framework named \system ~(a \underline{Deep} Neural Network-based \underline{I}ntrusion \underline{De}tector with \underline{A}ttack-sharing Loss). \system ~takes full advantage of deep learning to extract features and learn the classification boundary. Besides, we design a new loss function named {\em attack-sharing loss function} that   eliminates the bias towards the majority/benign class by moving the decision boundary towards the attack classes. To our best knowledge, this is the first work on imbalanced deep learning for intrusion detection.  Specifically, we make the following contributions.
     
First, we construct a deep feedforward network to learn intricate patterns of benign communications and malicious connections from the training data. To expedite the learning process on large data, we adapt a novel optimization algorithm that keeps track of an exponentially decaying average of the first-order and second-order moment of past gradients to dynamically adjust the learning rate.

Second, to address the class imbalance problem in intrusion detection, we design a new loss function named attack-sharing loss for our deep feedforward network. The attack-sharing loss function takes the discrepancy penalty of different types of mis-classification (e.g., mis-classifying attack types versus mis-classifying intrusion as normal) into consideration, so that the mis-classification of intrusions as benign receives more penalty than the mis-classification of attacks. 
It can be integrated with any deep neural network to mitigate the bias towards the majority class.

Last but not least, we launch an extensive set of experiments on three benchmark datasets. The comparison with 8 baseline approaches demonstrate the effectiveness of \system. In particular, \system ~produces the best detection accuracy on every dataset.

The rest of the paper is organized as follows. Section \ref{sc:pre} discusses the background information. Section \ref{sc:method} presents our the design of \system. Section \ref{sc:exp} shows the experiment results. Section \ref{sc:related} introduces the related work. Finally, Section \ref{sc:concl} concludes the paper.

\section{Background}
\label{sc:pre}
In this section, we introduce the background knowledge, including the concepts of deep neural network, intrusion attacks, and imbalanced classification.
\subsection{Deep Neural Network}
\label{sc:mlp}
Multi-layer perceptrons (MLPs), also known as deep feedforward network, is a network that consists of an input layer, multiple hidden layers, and an output layer. Each layer includes a certain number of neurons/units. The neurons in consecutive layers are connected by links with certain weights. 
Besides MLPs, other specialized architectures have been proposed in recent years. For example, convolutional networks are known for image processing, while recurrent neural networks are specialized at capturing long term dependencies \cite{goodfellow2016deep}.
Learning  the parameters (i.e., weights and bias) of a neural network is typically solved by using gradient descent. 
Back propagation provides an efficient way to calculate the gradients so as to optimize the weights associated with the connections.
Various optimization algorithms were proposed to acclerate the learning process, e.g., stochastic gradient descent (SGD) \cite{lecun1998gradient}, Nesterov Momentum \cite{mikolov2013distributed}, and Adam optimizer \cite{kingma2014adam}. 

\subsection{Intrusion Attacks}
Numerous types of network intrusion attacks make it intricate to design an effective detection method. Next, we briefly introduce five prevailing attacks that are investigated in this paper. 

\begin{itemize}
    \item {{\em Brute-Force} attack} is the most simple attack to gain illegal access to a site or server. The most common brute-force attack is the dictionary attack that cracks user passwords. There have been a few successful brute-force attacks. For example, in 2016, a massive brute-force attack against Alibaba Inc, a Chinese e-commerce giant, compromised 20.6 million accounts \cite{alibaba}. 

    \item {{\em Botnet} attack} exploits a number of Internet-connected devices (zombies) to carry out malicious and criminal tasks. As an example, a recent study \cite{echeverria2017star} of 6 million Twitter accounts reveals that 350,000 of them are zombies of the Star Wars botnet.  

    \item {{\em Probing} attack} scans a victim device in order to determine the vulnerabilities that can be exploited to compromise the system. It usually uses a network mapper (e.g., Nmap\footnote{\url{https://nmap.org/}}) to send TCP packets to discover vulnerable hosts and services. 

    \item {{\em DoS/DDoS} attack} overloads the target machine and prevents it from serving the intended users. 
    DDoS is different from DoS mainly in that it leverages multiple systems to exhaust the resources of the victim. The famous 2016 DDoS attack against Dyn DNS \cite{dyn} made it impossible for users to connect to Amazon, GitHub, Spotify, etc. 
    
    \item {{\em Infiltration} attack} leverages the vulnerability in particular software such as Adobe Acrobat Reader to execute a backdoor on the target computer. Once the attack is successful, the attacker can launch various types of attacks against the victim's network, including IP sweep and port scan. 
\end{itemize}

\subsection{Imbalanced Classification}
In general, the intrusion detection problem can be modeled as a classification problem in machine learning, by which the classification model outputs if the network system is intruded or not. 
In most real-world datasets, the data labels follow a long tail distribution, i.e., some specific classes are represented by a very small number of instances compared to other classes. In the scenario of intrusion detection, the data is also imbalanced, as the benign network behaviors dominate the collected dataset, while the intrusion events are rarely observed. 
To improve the overall accuracy, the imbalanced data forces the classification model to be biased toward the majority classes. 
This class imbalance problem renders poor accuracy on detecting intrusion attacks, as intrusion classes are under-represented. 
Several approaches \cite{zhou2006training} try to mitigate the negative effects of imbalanced data for general classification problems. 
One solution is to do {\em over-sampling} \cite{japkowicz2002class} or {\em under-sampling} \cite{kubat1997addressing}. In particular, over-sampling on the under-represented classes  duplicates these instances, while {\em under-sampling} \cite{kubat1997addressing} eliminates samples in the over-sized classes. However, over-sampling often leads to over-fitting and longer training time, and under-sampling degrades the overall accuracy since it discards potentially useful training instances.
Another solution is to make the loss function cost-sensitive by associating larger error penalty with under-represented classes \cite{khan2018cost}. However, in deep learning, such cost-sensitive loss function can make the loss of a minibatch highly sensitive to the label distribution. As the consequence, it leads to non-convergence of  the training process, and potentially inferiors the decision boundary of the classifier.

\nop{
\subsection{Convolutional Neural Network}
\label{sc:cnn}
Convolution is a special operation in deep neural network that applies a kernel function over the input data. Typically, the input is a multidimensional array of data, while the kernel is a multidimensional array of parameters that are tuned via the learning process. 
Any neural network that exploits convolution at any place is called a {\em convolutional neural network} \cite{krizhevsky2012imagenet}. 
A convolution layer consists of three stages, namely the {\em convolution stage} that performs linear combination of the input neurons, the {\em detector stage} that applies non-linear activation, and the {\em pooling stage} that summarizes the response over a neighborhood by using an aggregation function such as {\em max pooling} and {\em average pooling}.
Since it allows for parameter sharing and equivalence representations, convolutional neural network is more efficient to optimize, and is invariant to small variation in the input. 
To further improve training efficiency, strided convolution \cite{long2015fully} is proposed to sample a small fraction of output from the convolution layer.
}

\section{Our Approach}
\label{sc:method}
 In this section, we present the details of our intrusion detector, namely \system. In particular, we will discuss the model, loss function, and the optimization procedure of \system. 

\subsection{Model} 
\begin{figure*}[!hbtp]
	\centering
	\includegraphics[width=0.7\textwidth]{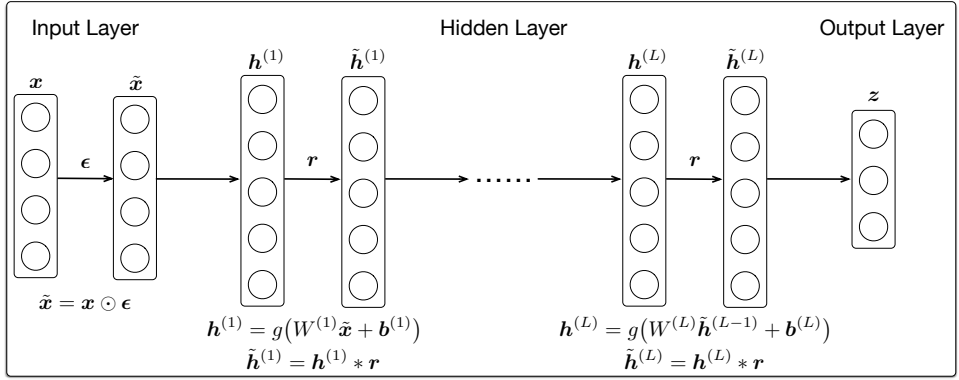}
	\vspace{-0.15in}
	\caption{\label{fig:structure}The model structure of \system~(Tildes indicate regularized layers)}
	\vspace{-0.2in}
\end{figure*}

At high level, \system~is built as a fully-connected neural network with one input layer, $L>1$ hidden layers, and one output layer. The input layer consists of $d$ units, each representing an input feature. 
The architecture of \system~is illustrated in Figure \ref{fig:structure}. Next, we explain the details of the model. 

We denote each input data point as $(\bm{x}, y)$, where $\bm{x}$ is the set of features, and $y$ is the label.  
The input layer of \system~consists of $d$ neurons that take input features.
For the sake of simplicity, for any instance $(\bm{x}, y)$, we assume $\bm{h}^{(0)}=\bm{x}$. 
We use $\bm{h}^{(\ell)}$ to denote the output of the $\ell$-th hidden layer ($1\leq \ell \leq L$). \Wendy{Does $\ell$ start from 0 or 1?} 
Let $n^{(\ell)}_i$ be the $i$-th unit of the $\ell$-th hidden layer. The output of $n^{(\ell)}_i$ is computed as:
\begin{equation}
	h^{(\ell)}_i = g(\bm{w}^{(\ell)}_i \tilde{\bm{h}}^{(\ell-1)} + b_i^{(\ell)}),
\end{equation} 
where $g(x)=max\{0, x\}$ is the {\em  rectified linear units} (ReLU) activation function, $\bm{w}^{(\ell)}_i$ is the $i$-th row in the weight matrix that connects the $(\ell-1)$-th and $\ell$-th layer, $b^{(\ell)}$ is the bias vector at the $\ell$-th layer, and $\tilde{\bm{h}}^{(\ell-1)}$ is the thinned output from the $(i-1)$-th layer by using dropout. In particular,
\begin{equation}
	\tilde{\bm{h}}^{(\ell-1)} = {\bm{h}}^{(\ell-1)} \ast \bm{r},
\end{equation}
where $\ast$ is the Hadamard product operator, $\bm{r}$ is a mask vector that specifies which units to be used. In particular, $\bm{r}$ consists of independent Bernoulli random variables, each of which has probability $p$ of being 1, i.e., $r_i \sim Bernoulli(p)$.

The output layer of the neural network includes $c$ {\em softmax} units, where $c$ is the number of classes. For any instance $(\bm{x}^{(i)}, y^{(i)})$, the predicted probability that it belongs to the $j$-th class $p_j^{(i)}$ is computed as
\begin{equation}
    \label{eq:p}
    p_j^{(i)} = softmax(\bm{z})_j =\frac{exp(z_j)}{\sum_{k=1}^{c} exp(z_k)},
\end{equation}
where $\bm{z}$ is a vector of linear activations of the output layer.

 \subsection{Attack-sharing Loss Function}
 Most modern neural networks use {\em cross-entropy loss} $J_{CE}$ to describe the discrepancy between the ground-truth labels and the model predictions. In particular, the loss $J_{CE}$ is calculated as: 
 \begin{align}
 \vspace{-0.1in}
  \label{eq:loss}
 J_{CE}(\bm{\theta}) & = \mathbb{E}_{(\bm{x}^{(i)},y^{(i)})\sim \hat{p}_{data}} L(f(\bm{x}^{(i)};\bm{\theta}), y^{(i)}) \nonumber \\
  & = -\mathbb{E}_{(\bm{x}^{(i)},y^{(i)})\sim \hat{p}_{data}} \log p(y^{(i)}|\bm{x}^{(i)};\bm{\theta}) \nonumber \\
  & = -\frac{1}{N} \sum_{i=1}^N \sum_{j=1}^c \mathbf{I}(y^{(i)},j)\log p_j^{(i)},
  \vspace{-0.1in}
 \end{align}
 where $\bm{\theta}$ consists of the weight matrix between consecutive layers in the neural network, $\hat{p}_{data}$ is the empirical data distribution in the training set, $p(y^{(i)}|\bm{x}^{(i)};\bm{\theta})$ is the probability that the neural network correctly classifies the input $\bm{x}^{(i)}$, $N$ is the number of training samples, $c$ is the number of classes, and $\mathbf{I}$ is the indicator function s.t.
 \begin{equation}
\mathbf{I}(a, b) = \begin{cases}
1 &\text{if } $a=b$\\
0 &\text{otherwise.}
\end{cases}
\end{equation}
The parameters $\bm{\theta}$ in the network are optimized so as to minimize $J_{CE}(\bm{\theta})$ and obtain the desired classification accuracy. 

One weakness of the cross-entropy loss function is that it does not take the type of mis-classification into consideration, and thus penalizes the classification error for all classes equally. There are two types of mis-classification for the intrusion detection system:
\begin{itemize}
    \item {\em Intrusion mis-classification}: an intrusion attack is mis-classified as benign event; and 
    \item{\em Attack mis-classification}: an intrustion attack of type A (e.g., DoS attack) is mis-classified as an intrution attack of type B (e.g., probing attack). 
\end{itemize}
In practice, the intrusion mis-classification should be penalized more than the attack mis-classification, as the attack mis-classification still triggers an alert to the IT security team, and enables the incident to be further inspected, whereas the intrusion mis-classification enables the attack incidents to bypass the security check and cause potentially critical damage.
Therefore, the intrusion mis-classification should have higher penalty than the attack mis-classification. \Wendy{I revised this part (starting from "One weakness of the basic...". Please double check if my understanding is correct.}


To address the issue of discrepancy penalty of different types of mis-classification, we improve the basic cross-entropy loss function. In particular, for any instance $(\bm{x}^{(i)}, y^{(i)})$, if it is a benign incident, $y^{(i)}=1$; otherwise, $y^{(i)} \in \{2, \dots, c\}$.
Inspired by \cite{shen2015deepcontour}, we design the {\em attack-sharing loss function}, $J_{AS}$, with an additional regularization term that penalizes intrusion mis-classification, i.e., the wrong estimation between the benign label and attack labels. In particular, we have
\begin{align}
\vspace{-0.1in}
\label{eq:as}
    J_{AS} & = J_{CE} -\frac{1}{N} \sum_{i=1}^N \lambda \big(\mathbf{I}(y^{(i)},1)\log p_1^{(i)} \nonumber \\
    & + \sum_{j=2}^c \mathbf{I}(y^{(i)},j)\log (1 - p_1^{(i)})\big),
    \vspace{-0.1in}
\end{align}
where $\lambda$ is a control parameter. When $\lambda$ is small, $J_{AS}$ is similar to the vanilla cross-entropy loss unction; when $\lambda$ is large, $J_{AS}$ tends to be an objective function for addressing the binary classification problem, benign versus attack. In our experiments, we set $\lambda=10$.
Compared with the basic cross-entropy loss, the attack-sharing loss function eliminates the bias towards the majority/benign class by moving the decision boundary towards the attack classes. It also respects the discrepancy penalty of different types of mis-classification. 

\subsection{Optimization Procedure} 
 In deep learning, the most widely-used optimization algorithm is {\em stochastic gradient descent} (SGD). In each round, it uses a minibatch of samples to estimate the gradient, and updates the parameters. Although simple, SGD suffers from slow asymptotic convergence, especially when there exist saddle points (i.e. points where one dimension slopes up and another slopes down) and plateaus (i.e., areas where the gradients keep stably high) in the parameter space. Due to the complicated nature of intrusion detection classification boundary, saddle points and plateaus widely exist. To expedite the learning process, we adapt the {\em Adam} optimizer, which adaptively updates the learning rate. 
In particular, we employ two variables, $\bm{s}$ and $\bm{r}$,  to store an exponentially decaying average of past gradients and squared gradients respectively. Initially, we set $\bm{s}=\bm{0}$ and $\bm{r}=\bm{0}$. In the $t$-th round of feedforward and backpropagation, we take a minibatch of $m$ samples from the training set, and calculate the stochastic gradient:
\begin{equation}
\label{eq:sgb}
\bm{g}_t \leftarrow \nabla_{\bm{\theta}_{t-1}} J(\bm{\theta}_{t-1}).
\end{equation}
Next, we update the first-order moment and second-order moment:
\begin{equation}
\bm{s}_t = \rho_1 \bm{s}_{t-1} + (1-\rho_1) \bm{g}_t,
\end{equation}
\begin{equation}
\bm{r}_t = \rho_2 \bm{r}_{t-1} + (1-\rho_2) \bm{g}_t^2,
\end{equation}
where $\rho_1, \rho_2\in (0,1)$ are the hyperparameters that determine the decay rate.
We also perform bias corrections to both moments to account for their initialization at the origin:
\begin{equation}
  \bm{s}_t = \frac{\bm{s}_t}{1-\rho_1^t},
\end{equation}
\begin{equation}
  \bm{r}_t = \frac{\bm{r}_t}{1-\rho_2^t}.
\end{equation}
Finally, the parameters are updated as 
\begin{equation}
\label{eq:adam}
    \bm{\theta}_t = \bm{\theta}_{t-1} - \frac{\zeta \bm{s}_t}{\sqrt{\bm{r}_t} + \delta},
\end{equation}
where $\zeta$ is the step size, and $\delta$ is a small stabilization factor. By using Equation (\ref{eq:adam}), we make greater evolution in the more gently sloped directions of parameter space. This facilitates faster convergence compared with SGD.
Another attractive property of the {\em Adam} optimizer is that it is robust to the choice of hyperparameters.

\section{Experiments}
\label{sc:exp}
\subsection{Dataset}
In our experiments, we use three datasets, namely {\em KDD99}, {\em CICIDS17} and {\em CICIDS18} dataset. Next, we briefly introduce these datasets.

\begin{table}[!htbp]
\vspace{-0.15in}
    \centering
    \caption{Class distribution in KDD99 dataset}
    \label{tab:dist_99}
    \vspace{-0.1in}
    \begin{tabular}{|c|c|c|c|c|}
        \hline
        \multirow{2}{*}{Label} & \multicolumn{2}{|c|}{Training} & \multicolumn{2}{|c|}{Testing} \\\cline{2-5}
        & Number & Fraction & Number & Fraction \\\hline
        Benign & 972,781 & 19.86\% & 60,593 & 19.48\%  \\
        DoS & 3,883,390 & 79.28\% & 231,455 & 74.42\% \\
        Probing & 41,102 & 0.84\% & 4,166 & 1.34\% \\
        U2R & 52 & 0.01\% & 245 & 0.08\%\\
        R2L & 1,106 & 0.02\% & 14,570 & 4.68\%\\\hline
        Total & 4,898,431 & 100\% & 311,029 & 100\% \\\hline
    \end{tabular}
    \vspace{-0.1in}
\end{table}

\noindent{\bf KDD99 dataset} is built by Stolfo et al. \cite{stolfo2000cost}, as a part of the DARPA Intrusion Detection Evaluation Program. 
To prepare this dataset, a military network environment was deployed to acquire nine weeks of raw TCP dump data from a local-area network (LAN). The LAN was operated as if it were a typical U.S. Air Force LAN, except that it was hacked by a sequence of cyber attacks. 
In this dataset, each connection record is described by 41 features and 1 label. 
The features include information in three aspects, namely basic connection information (e.g., duration, protocol type (tcp, udp, icmp), number of wrong fragments, number of urgent packets, etc), content information (e.g., number of failed login attempts, number of shell prompts, number of operations on access control files, etc), and traffic information (e.g., number of connections to the host in the past two seconds, fraction of connections that have ``SYN'' errors, etc).
The attacks in the dataset fall into 4 categories, i.e., DoS, Probing, U2R (normal users illegally gain root access to the system), and R2L (remote attackers exploit some vulnerabilities to obtain local access to the host).
We show the distribution of these attacks in the training and testing set in Table \ref{tab:dist_99}. 

\begin{table}[!htbp]
\vspace{-0.15in}
    \centering
    \caption{Class distribution in CICIDS17 dataset}
    \label{tab:dist_17}
    \vspace{-0.1in}
    \begin{tabular}{|c|c|c|c|c|}
        \hline
        \multirow{2}{*}{Label} & \multicolumn{2}{|c|}{Training} & \multicolumn{2}{|c|}{Testing} \\\cline{2-5}
        & Number & Fraction & Number & Fraction \\\hline
        Benign & 1,911,674 & 81.57\% & 361,399 & 74.84\%  \\
        DoS & 170,508 & 7.27\% & 82,151 & 17.01\% \\
        DDoS & 101,024 & 4.31\% & 27,003 & 5.59\% \\
        Brute-Force & 10,494 & 0.45\% & 3,341 & 0.69\%\\
        Infiltration & 149,934 & 6.40\% & 9,032 & 1.87\%\\\hline
        Total & 2,343,634 & 100\% & 482,926 & 100\% \\\hline
    \end{tabular}
    \vspace{-0.1in}
\end{table}

\noindent{\bf CICIDS17 dataset} is collected by Canadian Institute for Cybersecurity and is publicly available\footnote{\url{https://www.unb.ca/cic/datasets/ids-2017.html}}.
Two networks, namely the attack network and victim network were constructed. Each network is a infrastructure that consists of routers, switches and a set of PCs running most of the common operating systems. In total, there are 2.83 million network connection instances, where each instance is described by 81 features. The features are similar to those of the KDD99 dataset dataset, but include more statistical information. We omit the details due to the space limit.
We manually split the dataset into a training set and a testing set with a $5:1$ size ratio.
The launched attacks include DoS, DDoS, Infiltration and Brute-Force attacks. 
We show the distribution of these attacks in the training and testing set in Table \ref{tab:dist_17}. 

\begin{table}[!htbp]
\vspace{-0.15in}
    \centering
    \caption{Class distribution in CICIDS18 dataset}
    \label{tab:dist_18}
    \vspace{-0.1in}
    \begin{tabular}{|c|c|c|c|c|}
        \hline
        \multirow{2}{*}{Label} & \multicolumn{2}{|c|}{Training} & \multicolumn{2}{|c|}{Testing} \\\cline{2-5}
        & Number & Fraction & Number & Fraction \\\hline
        Benign & 4,197,451 & 82.62\% & 814,704 & 76.62\%  \\
        DoS & 517,691 & 10.19\% & 158,098 & 14.87\% \\
        Infilteration & 131,844 & 2.60\% & 38,787 & 3.65\%\\
        Botnet & 233,085 & 4.59\% & 51,753 & 4.87\%\\\hline
        Total & 5,080,071 & 100\% & 1,063,342 & 100\% \\\hline
    \end{tabular}
    \vspace{-0.1in}
\end{table}

\noindent{\bf CICIDS18 dataset} is also constructed by Canadian Institute for Cybersecurity and is publicly available\footnote{\url{https://www.unb.ca/cic/datasets/ids-2018.html}}. 
To simulate a real-world network, a common LAN network topology is implemented on the AWS computing platform.
The attacking infrastructure includes 50 machines and the victim organization has 5 departments and includes 420 machines and 30 servers. 
The dataset includes 6.3 million network connection instances, and each instance has 77 features. Again, we manually split the dataset into a training set and a testing set with a $5:1$ size ratio. 
The attacks in this dataset includes DoS, Infiltration and Botnet.
The class distribution of these attacks can be found in Table \ref{tab:dist_18}. 

\begin{table}[!htbp]
\vspace{-0.15in}
    \centering
    \caption{Summary of the datasets}
    \label{tb:data_sum}
    \vspace{-0.1in}
    \begin{tabular}{|c|c|c|c|c|c|}
        \hline
        Dataset & \# of & Training Size & Testing Size & \# of  & $\Omega_{imb}$  \\
        & Features & & & Classes &  \\\hline
        KDD99 & 41 & 4,898,431 & 311,029 & 5 & 2.96 \\\hline
        CICIDS17 & 81 & 2,343,634 & 482,926 & 5 & 3.08 \\\hline
        CICIDS18 & 77 & 5,080,071 & 1,063,342 & 4 & 2.31 \\\hline
    \end{tabular}
    \vspace{-0.1in}
\end{table}
We summarize the characteristics of these three datasets in Table \ref{tb:data_sum}. To evaluate the level of class imbalance in each dataset, we also report the {\em class imbalance measure} $\Omega_{imb}$ \cite{dong2018imbalanced} in the training set, which is defined as 
\begin{equation}
    \Omega_{imb} = \frac{\sum_{i=1}^{c} n_{max}-n_i}{n},
\end{equation}
where $n$ denotes the number of instances in the datast, $n_i$ denotes the number of instances that belong to the $i$-th class, and $n_{max} = max_{i=1}^c n_i$. Intuitively, $\Omega_{imb}$ measures the minimum percentage count of data samples required over all classes in order to form an overall balanced/uniform distribution. A larger $\Omega_{imb}$ value indicates higher level of class imbalance.

\begin{table*}[!htbp]
    \centering
    \caption{Detection accuracy comparison between \system ~and the baselines on the KDD99 dataset}
    \label{tab:accu_99}
    \vspace{-0.1in}
    \begin{tabular}{|c|c|c|c|c|c|c|c|c|c|c|c|}
        \hline
        \multirow{2}{*}{Classifier} & \multicolumn{2}{|c|}{Benign} & \multicolumn{2}{|c|}{DoS} & \multicolumn{2}{|c|}{Probe} & \multicolumn{2}{|c|}{U2R} & \multicolumn{2}{|c|}{R2L} &
        \multirow{2}{*}{CBA} \\\cline{2-11}
         & Pre & Rec & Pre & Rec & Pre & Rec & Pre & Rec & Pre & Rec & \\\hline 
         SVM  &  {30.2} & 70.04 & 96.66 & 98.55 & 13.68 & 27.71 & 21.87 & 6.76 & {84.4} & 5.91 & 41.23 \\\hline
         KNN  &  23.51 & 55.16 & 90.97 & 100 & 100 & 26.91 & 0 & 0 & 0 & 0 & 16.41 \\\hline
         DT  &  20.64 & 48.23 & 88.9 & 99.87 & 0 & 0 & 0 & 0 & 0 & 0 & 29.49 \\\hline
         MLP+CE  & 0 & 0 & 70.92 & {100} & 5.36 & 15.33 & 0 & 0 & 72.29 & 6.4 & 24.27 \\\hline
         MLP+OverSampling \cite{japkowicz2002class} & 27.15 & 11.68 & {95.5} & 83.14 & 5.64 & {74.64} & 11.83 & 1.97 & 62.79 & 19.15 & 38.12 \\\hline
         MLP+UnderSampling \cite{kubat1997addressing} & 19.21 & 45.95 & {95.52} & 83.57 & 9.43 & 33.87 & {20.83} & 8.09 & 68.3 & {29.78} & 40.25 \\\hline
         Cost-Sensitive \cite{khan2018cost} & 33.15 & 10.59 & 0 & 0 & 4.82 & 8.27 & 2.55 & 73.76 & 0 & 0 & 18.26 \\\hline
         CNN \cite{chowdhury2017few} & 20.50 & 64.79 & 94.88 & 82.81 & 10.53 & 27.27 & 7.14 & 1.3 & 60.87 & 4.67 & 36.17 \\\hline
         \system (Our approach) & 19.46 & {85.82} & 93.03 & 85.68 & 7.14 & 47.87 & 0 & 0 & 62.09 & 8.89 & {\bf 45.31}  \\\hline
    \end{tabular}
    \vspace{-0.1in}
\end{table*}

\begin{table*}[!htbp]
    \centering
    \caption{Detection accuracy comparison between \system ~and the baselines on the CICIDS17 dataset}
    \label{tab:accu_17}
    \vspace{-0.1in}
    \begin{tabular}{|c|c|c|c|c|c|c|c|c|c|c|c|}
        \hline
        \multirow{2}{*}{Classifier} & \multicolumn{2}{|c|}{Benign} &
        \multicolumn{2}{|c|}{DoS} &
        \multicolumn{2}{|c|}{DDoS} & \multicolumn{2}{|c|}{Brute-Force} & \multicolumn{2}{|c|}{Infiltration} & 
        \multirow{2}{*}{CBA} \\\cline{3-12}
          & Pre & Rec & Pre & Rec & Pre & Rec & Pre & Rec & Pre & Rec & \\\hline 
         SVM & 86.42 & 76.38 & {96.58} & 53.74 & {92.62} & 16.03 & 0 & 0 & 7.27 & {86.18} & 46.47 \\\hline
         KNN & 91.92 & 85.05 & 75.88 & 48.22 & 72.56 & 86.23 & 0 & 0 & 10.92 & 84.75 & 60.85 \\\hline
         DT & 66.51 & {100} & 0 & 0 & 0 & 0 & 0 & 0 & 0 & 0 & 20 \\\hline
         MLP+CE & 87.04 & 90.76 & 74.12 & {63.69} & 74.73 & 79.53 & 7.37 & 4.8 & 28.03 & 61.54 & 60.06 \\\hline
         MLP+OverSampling \cite{japkowicz2002class} & 86.03 & 95.05 & 80.14 & 52.5 & 56.68 & 76.06 & 3.65 & 1.63 & {28.18} & 53.62 & 55.45 \\\hline
         MLP+UnderSampling \cite{kubat1997addressing} & 86.88 & 54.9 & 50.91 & 59.31 & 26.13 & 11.32 & 7.17 & {27.39} & 13.8 & 58.03 & 42.19  \\\hline
         Cost-Sensitive \cite{khan2018cost} & 61.58 & 61.17 & 17.69 & 28.09 & 0 & 0 & 0 & 0 & 0 & 0 & 17.85 \\\hline
         CNN \cite{chowdhury2017few} & 0 & 0 & 23.42 & 96.04 & 0 & 0 & 8.07 & 11.07 & 0 & 0 & 21.42 \\\hline
         \system (Our approach) & 88.5 & 94.06 & 88.77 & 62.97 & 76.31 & 83.19 & {8.29} & 4.1 & 26.46 & 64.53 & {\bf 61.77} \\\hline
    \end{tabular}
    \vspace{-0.1in}
\end{table*}

\begin{table*}[!htbp]
    \centering
    \caption{Detection accuracy comparison between \system ~and the baselines on the CICIDS18 dataset}
    \label{tab:accu_18}
    \vspace{-0.1in}
    \begin{tabular}{|c|c|c|c|c|c|c|c|c|c|}
        \hline
        \multirow{2}{*}{Classifier} & \multicolumn{2}{|c|}{Benign} & \multicolumn{2}{|c|}{DoS} & \multicolumn{2}{|c|}{Infiltration} & \multicolumn{2}{|c|}{Botnet} &
        \multirow{2}{*}{CBA} \\\cline{2-9}
         & Pre & Rec & Pre & Rec & Pre & Rec & Pre & Rec & \\\hline 
         SVM & {76.59} & 42.59 & 44.97 & 26.23 & {26.57} & 20.64 & 13.70 & 29.14 & 29.65 \\\hline
         KNN & 75.46 & 30.05 & 31.2 & 28.05 & 11.53 & 6.45 & {17.32} & 28.14 & 23.18\\\hline
         DT & 64.77 & {100.0} & 0 & 0 & 0 & 0 & 0 & 0 & 25.00  \\\hline
         MLP+CE & 67.42 & 71.73 & 31.92 & 45.23 & 0 & 0 & 18.52 & 0.42 & 29.35 \\\hline
         Cost-Sensitive \cite{khan2018cost} & 54.52 & 19.6 & 0 & 0 & 5.44 & {71.42} & 10.38 & 6.54 & 24.39 \\\hline
         CNN \cite{chowdhury2017few} & 0 & 0 & 22.44 & 98.52 & 7.08 & 8.19 & 0 & 0 & 26.68 \\\hline
         \system (Our approach) & 70.96 & 39.7 & 45.47 & 49.94 & 6.88 & 21.0 & 11.86 & {33.64} & {\bf 36.07} \\\hline
    \end{tabular}
    \vspace{-0.1in}
\end{table*}

\subsection{Setup}
We implement \system ~in Python by using the {\em tensorflow} framework. 
In our neural network, we include 100 units in each hidden layer. We try various number of hidden layers. Our results suggest that the performance is reasonably good as long as the network consists of at least 6 layers. In the following section, we only show the results with 10 hidden layers.
The keep probability of each dropout layer is 0.8. The source code is available in a public repository\footnote{\url{https://github.com/bxdong7/MLP-D}}. 
In our experiments, we group 100 consecutive network connection instances as a training sample. We set the learning rate as $10^{-4}$, the minibatch size as 128, and the number of epoches as 10. We use cross-validation to avoid overfitting.
We run \system ~on a workstation with 1 NVIDIA RTX 2080 Ti GPU and 1 Intel i7-7700K @ 4.2GHz CPU. On average, the training process halts within 3 hours.

\subsection{Baseline}
We compare \system ~ with the following baseline approaches.
    \begin{itemize}
        \item SVM (support vector machine) with 100 iterations
        \item KNN (k-nearest neighbor) with 5 neighbors and minkowski distance 
        \item DT (decision tree) with 10 layers at most
        \item MLP+CE (multi-layer perceptron with cross-entropy loss function (Formula (\ref{eq:loss})))
        \item MLP+OverSampling \cite{japkowicz2002class}: In this approach, the training data is transformed to balanced class distribution by applying over-sampling.
        \item MLP+UnderSampling \cite{kubat1997addressing}: In this approach, the training data is transformed to balanced class distribution by applying over-sampling.
        \item Cost-Sensitive \cite{khan2018cost}: This approach uses cost-sensitve loss function to train the neural network. We follow \cite{khan2018cost} to set up the cost matrix.
        \item CNN \cite{chowdhury2017few}: This approach classifies the network connections with 2 convolution layers, 2 maxpooling layers and 6 fully-connected layers. 
    \end{itemize}
We implement these machine learning-based baselines by using the {\em scikit-learn} library, 
and the deep learning-based baselines by using the {\em tensorflow} framework. 
We also implement a baseline approach based on recurrent neural network with LSTM units \cite{hochreiter1997long}. However, the training process does not complete within 5 days. Therefore, we do not include the results in this section.

\subsection{Evaluation Metrics}
In this paper, we are mostly interested in evaluating the detection accuracy. In the experiments, we collect the following statistics: TP (true positive), FP (false positive), TN (true negative) and FN (false negative).
\nop{
\begin{itemize}
    \item True Positive (TP): number of true attacks that are successfully identified.
    \item False Positive (FP): number of false alerts signified by the detection approach.
    \item True Negative (TN): number of benign connections that are correctly classified.
    \item False Negative (FN): number of true attacks that are missed by the detection approach.
\end{itemize}
}
We measure the following the {\em precision} and {\em recall} for each class. In particular, $precision = \frac{TP}{TP+FP}$ measures the fraction of alerts indicated by the detection approach that are correct, and $recall = \frac{TP}{TP+FN}$ measures the fraction of attacks that are successfully identified by the detection approach.
\nop{
\begin{itemize}
    \item Precision: the fraction of alerts indicated by the detection approach that are correct, i.e., 
    \[precision = \frac{TP}{TP+FP}.\]
    \item Recall: the fraction of attacks that are successfully identified by the detection approach, i.e., 
    \[recall = \frac{TP}{TP+FN}.\]
\end{itemize}
}

In addition, we measure the overall class-balanced accuracy (CBA) \cite{dong2018imbalanced} for all classes. CBA is calculated as the average recall for all classes. By taking all classes equally important, it avoids inflated performance estimates on imbalanced datasets. If the classifier performs equally well on every class, this term is equivalent to the conventional accuracy (i.e., the number of correct predictions divided by the total number of predictions). On the contrary, if the model only performs well on the majority class, CBA drops to $\frac{1}{c}$.
Therefore, CBA is an effective metric in measuring the accuracy of a classifier for imbalanced dataset.


\nop{
\begin{figure}[!hbtp]
    \centering
    \includegraphics[width=0.45\textwidth]{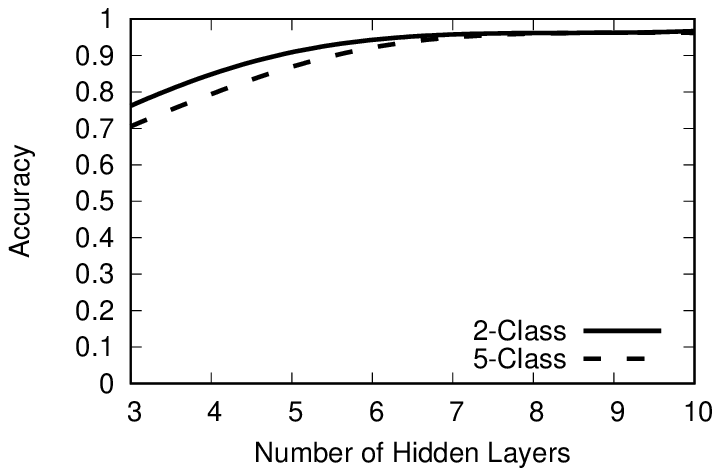}
    \caption{Accuracy v.s. number of hidden layers}
    \label{fig:accuracy_vs_layers}
\end{figure}
}

\vspace{-0.05in}
\subsection{Evaluation of \system}

\noindent{\bf KDD99 Dataset.} 
We present the classification accuracy of \system ~ and the baseline approaches on the KDD99 dataset in Table \ref{tab:accu_99}. First, we observe that \system ~yields the highest CBA. This demonstrates that \system ~is effective in detecting intrusion attack incidents from imbalanced dataset. 
Second, KNN, DT and MLP+CE only focus on the majority classes, and produces unsatisfactory performance on the minority classes. For instance, KNN and DT fail to catch any U2R and R2L attack instance. 
In contrast, the cost-sensitive classifier associates too much cost for the U2R and R2L classes, as they are extremely under-represented. This makes the classifier lean too much toward these classes, and  performs poorly on the KDD99 dataset.
Over-sampling and under-sampling mitigate the side-effect of the class imbalance problem, but the improvement is not comparable with \system.

\noindent{\bf CICIDS17 Dataset.}
We report the accuracy of all the classifiers in Table \ref{tab:accu_17}. 
First, we observe that \system ~produces similar and satisfying precision and recall on every class, except for the {\em Brute-Force}. Consequently, \system ~yields the best CBA among all the classifiers. In Table \ref{tab:dist_17}, it is easy to see that {\em Brute-Force} is the most under-represented class. The {\em Brute-Force} attack instances only takes around 0.5\% of the dataset. Even though the attack-sharing loss function aims at dragging the decision boundary toward the attack classes, it does not help much with this class. 
Second, the performance of KNN and MLP+CE are close to that of \system. We further investigate the reason and find that a large fraction of the attack instances are present in both the training and testing set.
DT simply labels every test instance as benign connection. Similarly, CNN \cite{chowdhury2017few} almost recognizes every connection as DoS attack. The cost-sensitive classifier only focuses on the benign and DoS class. Naturally, the CBA of these three baselines are low. 


\noindent{\bf CICIDS18 Dataset.}
In Table \ref{tab:accu_18}, we compare the accuracy of \system ~with the baselines on the CICIDS18 dataset.
\system ~performs the second best in addressing the class imbalance problem. It has similar recall on every class. Again, it shows the effectiveness of the class-sharing loss function. 
No baseline approach produces a CBA higher than 30\%. 
Again, the cost-sensitive baseline concentrates all the attention to the most under-represented class, i.e., infilteration, and neglects the other classes. DT simply recognizes every testing instance as benign. 

\nop{
\begin{table}[!hbtp]
    \centering
    \caption{2-class detection accuracy comparison}
    \label{tab:2class}
    \begin{tabular}{|c|c|c|c|c|c|c|c|c|c|}
        \hline
        \multirow{2}{*}{Classifier} & 
        \multicolumn{3}{|c|}{KDD99} &
        \multicolumn{3}{|c|}{CICIDS17} &
        \multicolumn{3}{|c|}{CICIDS18} \\\cline{2-10}
        & Acc & Pre & Rec & Acc & Pre & Rec & Acc & Pre & Rec \\\hline
        SVM & 85.89 & 65.75 & 95.92 \\\hline
        KNN & 87.54 & 68.65 & 96.14 \\\hline
        DT & 96.03 & 89.99 & 95.41 \\\hline
        NB \cite{belouch2018performance} & 74.19 & N/A & 92.16 \\\hline
        MLP & 80.10 & 56.83 & {\bf 98.77} \\\hline
        CNN + SVM \cite{chowdhury2017few} & 95.51 & N/A & N/A \\\hline
        LSTM \cite{kim2017effective} & 95.92 & {\bf 96.75} & N/A \\\hline
        \system (Our approach) & {\bf 96.47} & 91.50 & 95.33 \\\hline
    \end{tabular}
    \vspace{-0.1in}
\end{table}
}

\subsection{Insights}
On all datasets, \system ~delivers the best CBA. The CBA of \system ~is always better and can as twice high as that of MLP+CE. Compared with the sampling-based approaches, the accuracy of \system ~on all classes is more balanced. \system ~significantly outperforms the cost-sensitive classifier \cite{khan2018cost}, which only focuses on a few classes. CNN \cite{chowdhury2017few} is inferior to \system~ in classification accuracy mainly because network events that are close in time do not exhibit similar behaviors. All these observations demonstrate the effectiveness of our attack-sharing loss function in dealing with class imbalance in intrusion detection dataset.

However, we must acknowledge the weakness of \system. It does not concentrate sufficiently on the extremely under-represented classes. For example, the U2R class in the KDD99 dataset and the Brute-Force class in the CICIDS17 dataset are rarely detected by \system. Both classes take no more than 1\% in the training and testing set. The reason is that the attack-sharing loss only pulls the decision boundary towards the attack classes. However, the tow direction is still biased towards the majority attack classes. In specific, the regularization term in Formula (\ref{eq:as}) does not differentiate different types of attacks. In order to minimize the $J_{AS}$ loss, the classifier tends to classify every attack instance as a majority attack class. 
This limitation makes \system ~more suitable for the scenarios where the benign instances dominate the dataset, and different types of attacks are balanced. In other words, \system ~works well when the imbalance only exists between the benign class and attack classes. 
\section{Related Work}
\label{sc:related}
\subsection{Intrusion Detection based on Deep Learning}
\cite{almseidin2017evaluation,biswas2018intrusion} conducted a thorough comparison of a variety of classifiers, including J48 tree, Naive Bayes, decision tree, support vector machine (SVM) and k-nearest neighbor (KNN), on the accuracy of intrusion detection.
To reduce the generalization error of the classifiers and avoid overfitting, Rigaki et al. 
Javaid et al. \cite{javaid2016deep} developed a novel intrusion detection approach based on {\em self-taught learning}. It is a deep neural network that consists of two stages. In the first stage, a new feature representation of the input data is learned from a sparse auto-encoder. After that, the new features are taken by a soft-max regression for classification. The experiment results demonstrate the effectiveness of the new feature representation in improving classification accuracy. 
Chowdhury et al. \cite{chowdhury2017few} applied 'few-shot learning' to improve the detection accuracy. In specific, they first train a convolutional neural network and then extract features from various layers. Those features are fed into various classifiers such as SVM or 1-nearest neighbor (1-NN). Experimental results suggest that the class-wise accuracy can reach 65.8\%. However, it is not clear what is the rationale behind the choice of SVM and 1-NN over a softmax activation in the last layer of a deep neural network. 
Kitsune \cite{mirsky2018kitsune} is the most recent work that detects network intrusion attacks with deep neural networks. It applies an ensemble of autoencoders to learn the identify function of the original data distribution. For any new instance, its anomaly score is calculated based on the distance between the autoencoders' output and its feature values. 
However, we argue that such a design only employs deep learning to discover inherent/generic features in network connections, but fails to take advantage of its capacity to learn complex classification functions. 

\subsection{Anomaly Detection based on Deep Learning}
Quite a few work concentrate on applying deep neural networks to detect abnormal behaviors that are not essentially intrusion attacks.
Zhang et al. \cite{zhang2016automated} propose to detect IT system failures from system log files. Clustering analysis is used to extract frequent patterns in the log files. Then the log files are embedded by counting the number of occurrences of these patterns. To cope with the lack of labeled data, a LSTM is trained to capture the long-range dependency across sequences.
Du et al. \cite{du2017deeplog} also applied LSTM to detect anomalies and diagnoze failures from system logs. Different from previous work, they aim at abnormal execution paths to improve the confidence and help with further investigation. 
Kiran et al. \cite{kiran2018overview} provide a comprehensive survey on the application of deep learning over anomaly detection in videos. 
Recently, Chalapathy et al. \cite{chalapathy2018anomaly} show that it is advantageous to directly train deep networks to extract progressively rich representation of data with the classification objective, rather than utilizing a hybrid approach where deep features are firstly learned by using an autoencoder and then fed into a separate anomaly detection method like SVM. 
Our work is consistent with this proposition, and thus produces higher accuracy compared with traditional hybrid approaches.

\section{Conclusion}
\label{sc:concl}
In this paper, we design \system, a novel intrusion detection and classification system based on imbalanced deep learning. To deal with the imbalanced class problem, we design a new loss function named attack-sharing loss to eliminate the bias towards the majority/benign class. We also integrate \system ~with a new optimization algorithm to facilitate efficient training. Experimental results on three benchmark datasets show the superiority of \system ~compared to seven baseline approaches. 

In the future, we plan to extend this work from the following perspectives. First, we plan to apply the attack-sharing loss to recurrent neural networks that take the network context into consideration. We also plan to use generative adversarial networks to improve the detection accuracy.

\bibliographystyle{splncs04}
\bibliography{bib}
\end{document}